\documentclass[twocolumn,prl]{revtex4-2}
\usepackage{graphicx}
\usepackage{amssymb}
\usepackage{epstopdf}
\usepackage{amsmath,bm}
\usepackage{bm}
\usepackage{multirow}

\usepackage{isomath}
\usepackage{footnote}
\usepackage[bottom]{footmisc}
\usepackage{color}
\makeatletter

\@ifundefined{textcolor}{}
{
 \definecolor{BLACK}{gray}{0}
 \definecolor{WHITE}{gray}{1}
 \definecolor{RED}{rgb}{1,0,0}
 \definecolor{GREEN}{rgb}{0,1,0}
 \definecolor{BLUE}{rgb}{0,0,1}
 \definecolor{CYAN}{cmyk}{1,0,0,0}
 \definecolor{MAGENTA}{cmyk}{0,1,0,0}
 \definecolor{YELLOW}{cmyk}{0,0,1,0}
}

\usepackage[colorlinks,
            linkcolor=black,
            anchorcolor=black,
            citecolor=black
            ]{hyperref}
\usepackage{ulem}           
            
\makeatother

\DeclareMathAlphabet\mathbfcal{OMS}{cmsy}{b}{n}

\begin{document}

\title{Protected Topological Nodal Ring Semimetal in Graphene}
\author{Karyn Le Hur and Sariah Al Saati}
\affiliation{CPHT, CNRS, Institut Polytechnique de Paris, Route de Saclay, 91128 Palaiseau, France}

\begin{abstract}
Graphene is a two-dimensional Dirac semimetal showing interesting properties as a result of its dispersion relation with both quasiparticles and quasiholes or matter and anti-matter. We introduce a topological nodal ring semimetal in graphene with a quantized quantum Hall response, a robust one-dimensional chiral edge mode and a quadratic Fermi-liquid spectrum for the quasiparticles and quasiholes in the bulk. The bulk band degeneracy at the Fermi energy is protected through a $\mathbb{Z}_2$ symmetry related to the two spin polarizations of an electron and a ``double-orthogonality'' structure in the sublattice and spin quantum numbers of the two crossing eigenstates. The system may have applications in nano-electronics and in quantum mechanical entanglement applied to band theory. 
\end{abstract}
\maketitle

{\it Introduction.---} The semimetallic state in graphene is formed around its two inequivalent Dirac points within the Brillouin zone and is characterized by a linear energy spectrum and density of states \cite{RMPgraphene}. The relation with the two-dimensional Dirac equation has resulted in various applications such as the observation of the Klein paradox \cite{Kleinobservation}. Yet, the Coulomb interaction may lead to corrections on thermodynamics and transport such as a logarithmic renormalization of the velocity of the quasiparticles \cite{Vozmediano}. Each valley surrounding a Dirac point gives rise to a $\pm\pi$ Berry phase \cite{Berry} reflecting the pseudospin-$\frac{1}{2}$ structure of the Hamiltonian or the helicity, but macroscopically the system is described through a zero total Berry phase from band theory, such that the quantum Hall conductivity \cite{Thouless} is also zero \cite{RMPgraphene}. In the presence of a perpendicular magnetic field \cite{Klitzing}, the system can then develop a quantum Hall response \cite{QAH1,QAH2} which becomes fractional \cite{QAHfractional1,QAHfractional2,QAHreview} as a result of interactions \cite{Laughlin,Hallfrac}.  Graphene materials show great promise for nanoelectronics with recent developments on bilayers and trilayers with Bernal \cite{Geneva} and Moir\' e stacking \cite{EvaReview} including the observation of topological insulating and superconducting phases of matter \cite{Hallbilayer,MIT,Saito,trilayer}. In this Letter, we introduce a topological nodal ring semimetal in a graphene plane with a non-zero density of states similar to a two-dimensional Fermi liquid and a robust quantized quantum Hall conductivity. We elaborate on a precise implementation of the system and on its topological nature, its stability towards disorder and interaction effects.  

At the heart of our model to realize the topological semimetal in graphene is the presence of a two spins-$\frac{1}{2}$ matrix theory defining a two-particles ground state wavefunction at half-filling. The system may be visualized as one filled energy band with properties similar to the Haldane model \cite{Haldane} and the quantum anomalous Hall effect revealing a topological invariant and a chiral edge mode, which coexists with a nodal ring semimetal surrounding one Dirac point. In the bulk, the semimetal gives rise to a Fermi liquid with the effective mass of the quasiparticles or quasiholes fixed through the band structure. The proximity to the lowest energy band makes the semimetal topological. In Ref. \cite{HH}, we have introduced the possibility of a topological nodal ring semimetal in a bilayer system with specific $AA-BB$ stacking. Here, in a monolayer (one-plane) graphene the two flavors refer precisely to the two spin polarizations of an electron, protecting the formation of the topological semimetal through the $\mathbb{Z}_2$ symmetry corresponding to invert the two spin polarizations in the model. The system is rare as a quantized invariant is usually the hallmark of a Chern insulator or a topological superconductor \cite{QiZhang}. A Fermi surface may generally lead to corrections in (topological) transport properties \cite{Haldane2004,Kagome}. We emphasize here the recent interest to semimetals showing topological properties \cite{HH,YoungKane,SemimetalsReview,BernevigFelser,Hu,Burkov}.

{\it Model.---} We begin with the class of Hamiltonians in the momentum space, $H=\sum_{\bf k} \psi^{\dagger}({\bf k}) {\cal H}({\bf k})\psi({\bf k})$ with $\psi({\bf k})=(c_{A{\bf k}\uparrow},c_{B{\bf k}\uparrow},c_{A{\bf k}\downarrow},c_{B{\bf k}\downarrow})$ and
\begin{eqnarray}
\label{Hmodel}
{\cal H}({\bf k}) &=& ({d}_z({\bf k}) + M)\sigma_z\otimes \mathbb{I} + d_{x}({\bf k}) \sigma_x\otimes \mathbb{I} \\ \nonumber
&+& d_{y}({\bf k})\sigma_y\otimes \mathbb{I} + r\mathbb{I}\otimes s_x.
\end{eqnarray}
Eq. (\ref{Hmodel}) is generally applicable on lattices with two inequivalent sites, referring to two `sublattices', and hereafter we address the situation of the honeycomb lattice related to graphene. Similarly as in the Kane-Mele model for two-dimensional topological insulators \cite{KaneMele}, the Hamiltonian is written in terms of two sets of Pauli matrices: $\mathbfit{\sigma}$ acting on the Hilbert space $\{|A\rangle;|B\rangle\}$ associated to the two sublattices $A,B$ and $\bf{s}$ acting on $\{|+\rangle_z;|-\rangle_z\}$ linked to the two spin polarizations of an electron. We fix $\hbar=1$ or $h=2\pi$ such that ${\bf k}=(k_x,k_y)$ refers equally to a wave-vector or momentum. Here, $c^{\dagger}_{i{\bf k}\alpha}$ creates an electron, corresponding to sublattice $i=A,B$, with momentum ${\bf k}$ and spin polarization $\alpha=\uparrow,\downarrow$ associated to $|+\rangle_z,|-\rangle_z$. The $\mathbb{Z}_2$ symmetry identically corresponds to invert the role of $|+\rangle_z \leftrightarrow |-\rangle_z$. A related aspect of the model is the presence of an in-plane magnetic field to realize the topological phase(s). The term $r$ can be realized via a Zeeman effect and a magnetic field $B_x$ applied in the $x$ direction such that $r=-\frac{\gamma B_x}{2}=\mu_B B_x>0$ with $\gamma$ the gyromagnetic factor and $\mu_B$ the Bohr magneton.

The components $d_x({\bf k})$ and $d_y({\bf k})$ are similar as in the Haldane model ${\cal H}({\bf k})={\bf d}({\bf k})\cdot\mathbfit{\sigma}$ \cite{Haldane} and encode the kinetic term in graphene while ${d}_z({\bf k})$ represents the topological term producing a ``mass-inversion'' effect at the two Dirac points $K$ and $K'$ within the Brillouin zone \cite{Notations}. From standard definitions on the honeycomb lattice \cite{KarynReview}, we identify $d_x({\bf k})=-t\sum_{i=1}^3 \cos({\bf k}\cdot {\mathbfit{\delta}}_i)$, $d_y({\bf k})=-t\sum_{i=1}^3 \sin({\bf k}\cdot {\mathbfit{\delta}}_i)$ with $t>0$ and ${d}_z({\bf k})=2t_2\sum_{j} \sin({\bf k}\cdot {\bf b}_j)$. Here, $t$ represents the hopping amplitude between nearest-neighbouring $A$ and $B$ sites, $\mathbfit{\delta}_i$ define the three vectors linking these sites, $t_2$ is the amplitude describing the hopping term between second-nearest neighbouring sites. Here, ${\bf b}_j$ define the three vectors in a triangle loop formed with the three sites belonging to one sublattice $A$ or $B$ within one honeycomb unit cell \cite{KarynReview}. The term ${d}_z({\bf k})$ can be produced through a Peierls phase attached to the $t_2$ term such that $t_2=|t_2| e^{i\phi}$ with a maximum effect when $\phi=\pm\frac{\pi}{2}$. For $\phi=-\frac{\pi}{2}$, $d_z=\zeta\tilde{d}_z$ with $\tilde{d}_z=3\sqrt{3}t_2$ and $\zeta=\pm$ at the $K$ and $K'$ points respectively. Hereafter, we show one  protocol to realize such a specific term through circularly polarized light which can be implemented with current technology \cite{Hamburglight}. The term ${d}_z$ may also be realized in cold atoms in optical lattices \cite{Jotzu,Hamburg} or in light systems \cite{Review1,Review2}. The parameter $M$ refers to a modulated potential or a Semenoff mass term \cite{Semenoff} which will drive the bands-crossing effect at zero energy for the half-filled situation.
 
For our purpose, it is useful to draw an analogy with the Bloch sphere formalism which allows for a characterization of the global topological properties from the poles or equivalently from the Dirac points on the lattice \cite{KarynReview,C2}. The topological information being in fact transported in a thin cylinder from the equatorial plane to each pole. The two Dirac points $K$ and $K'$ are located at the north and south poles respectively. Introducing a small deviation ${\bf p}=(p_x,p_y)$ from each Dirac point, we have $d_x({\bf p})=v_F p_x=v_F|{\bf p}|\cos\tilde{\varphi}=|{\bf d}|\cos\varphi\sin\theta$, $d_y({\bf p})=\zeta v_F p_y=\zeta v_F|{\bf p}|\sin\tilde{\varphi}=|{\bf d}|\sin\varphi\sin\theta$ such that $d_x({\bf p})^2 + d_y({\bf p})^2 = v_F^2 |{\bf p}|^2$ with $\tilde{\varphi}$ being related to the azimuthal angle on the sphere $\tilde{\varphi}=\zeta\varphi$. The Fermi velocity defined as $v_F=\frac{3}{2}t a$, with $a$ the lattice spacing, is $300$ times smaller than the speed of light $c$ \cite{RMPgraphene}. Additionally, we identify $d_z({\bf p})=\zeta\tilde{d}_z+M=|{\bf d}|\cos\theta$ with $\theta$ the polar angle in spherical coordinates and we can equally write $|{\bf d}|^2={\bf d}\cdot {\bf d}=d_x({\bf p})^2+d_y({\bf p})^2+(\zeta\tilde{d}_z+M)^2$. For $r=0=M$, each Bloch sphere characterizing one spin polarization gives rise to a radial magnetic field producing one Dirac monopole or Skyrmion related to the Haldane model \cite{Haldane}. As long as $M<\tilde{d}_z$ and $r=0$, the topological properties remain identical and for $M=\tilde{d}_z$ the topological charge leaks out from the sphere. Therefore, below we assume that $M<\tilde{d}_z$. To classify the topological properties, it is judicious to introduce the pseudo-spin eigenstates $|\psi_+\rangle$ and $|\psi_-\rangle$ corresponding to energies $\pm |{\bf d}|$ and $|+\rangle_x$, $|-\rangle_x$ corresponding to the spin polarizations $s_x=\pm 1$ such that $|\pm\rangle_x = \frac{1}{\sqrt{2}}(\pm |+\rangle_z + | - \rangle_z)$ (which maybe redefined modulo a global phase). We will then specify the forms of the eigenstates close to the Dirac points or poles completed by numerical calculations. 

\begin{table}[!hbt]
\begin{tabular}{||c|c|c||}
    \hline
    & band 2 & band 3 \\
    \hline
    \multirow{2}{*}{$K$} & $E_2(K) = r - \left|\bm{d}(K)\right| $ &  $E_3(K) = - r + \left|\bm{d}(K)\right| $   \\
    & $|\psi_2(K)\rangle = |\psi_{-}\rangle\otimes |+\rangle_x$ & $|\psi_3(K)\rangle = |\psi_{+}\rangle \otimes |-\rangle_x$  \\
    \hline
    \multirow{2}{*}{$K^\prime$}  &  $E_2(K^\prime) = - r + \left|\bm{d}(K^\prime)\right| $ & $E_3(K^\prime) = r - \left|\bm{d}(K)\right| $\\
    & $|\psi_2(K^\prime)\rangle = |\psi_{+}\rangle\otimes |-\rangle_x$  & $|\psi_3(K^\prime)\rangle = |\psi_{-}\rangle \otimes |+\rangle_x$ \\
    \hline
    \end{tabular}
    \caption{Energies and corresponding eigenstates of band 2 and band 3 at the Dirac points.}
\label{tab:statesDiracPoints}
\vskip -0.5cm
\end{table}

{\it Semimetal Characteristics.---} One key objective of this Letter is to show that Eq. (\ref{Hmodel}) allows for a simple physical and mathematical understanding for the existence of such a protected topological nodal ring semimetal. To fix the parameters accordingly, it is fruitful to study ${\cal H}({\bf k})^2$, the square of the Hamiltonian in the vicinity of $K$ and $K'$
\begin{equation}
\label{H2}
{\cal H}({\bf k})^2 = (|{\bf d}({\bf k})|^2 +r^2)\mathbb{I}\otimes \mathbb{I} +2r{\bf d}({\bf k})\cdot \mathbfit{\sigma}\otimes s_x.
\end{equation}
The system generally develops four energy bands. The lowest and top energy levels correspond to the eigenstates $|\psi_1\rangle=|\psi_-\rangle \otimes |-\rangle_x$ and $|\psi_4\rangle=|\psi_+\rangle \otimes |+\rangle_x$ assuming $r>0$. Then, $2r{\bf d}\cdot \mathbfit{\sigma} \otimes s_x=2r{\bf d}\cdot \mathbfit{\sigma}$ when acting on these two energy states. Therefore, the lowest and uppest energies in the spectrum satisfy $E^2 = (r+|{\bf d}|)^2$ with respectively $E_1=-(r+|{\bf d}|)$ and $E_4 = (r+|{\bf d}|)$. It is perhaps judicious to emphasize that at the Dirac points we have $d_x=d_y=0$ such that $E_1(K)=-(r+\tilde{d}_z+M)$ and $E_1(K')=-(r+\tilde{d}_z-M)$. Similarly, we identify $E_4(K)=r+\tilde{d}_z+M$ and $E_4(K')=r+\tilde{d}_z-M$. 

The two middle or intermediate bands correspond to the two eigenstates $|\psi_-\rangle\otimes |+\rangle_x$ and $|\psi_+\rangle \otimes |-\rangle_x$ such that $2r{\bf d}\cdot \mathbfit{\sigma} \otimes s_x=-2r{\bf d}\cdot \mathbfit{\sigma}$. These two energy bands are then described through $E^2=(-r+|{\bf d}|)^2$. To observe a nodal ring semimetal at half-filling, this requires these two bands to meet at zero energy such that $E^2=0$. If these quantum states join sufficiently close to one Dirac point, for instance $K'$, this gives rise to the equality \cite{Remark}
\begin{equation}
\label{radius}
v_F^2 |{\bf p}|^2 = r^2 - (\tilde{d}_z-M)^2
\end{equation}
assuming that $r>(\tilde{d}_z-M)$.
From the organization of energy bands defined such that $E_1<E_2<E_3<E_4$, since $E_2(K)<E_3(K)$ with $E_2(K)=r-|{\bf d}(K)|=r-\tilde{d}_z-M$ and $E_3(K)=-r+|{\bf d}(K)|=-r+\tilde{d}_z+M$ then this also implies that $r<d_z+M$. The formation of a (topological) nodal ring semimetal then leads to
\begin{equation}
\label{prerequisite}
\tilde{d}_z-M<r<\tilde{d}_z+M,
\end{equation}
which interestingly provides a correspondence towards the analysis on two interacting Bloch spheres \cite{HH}.  A key property of the model is the phenomenon of inversion of the forms of eigenstates related to bands $2$ and $3$ at $K$ and $K'$ as a result of the Semenoff term $M$ and $r$, which is summarized in the Table for clarity sake. We emphasize that $|{\bf d}(K)|=\tilde{d}_z+M$ and $|{\bf d}(K')|=\tilde{d}_z-M$.
The $|\psi_-\rangle$ $(|\psi_+\rangle)$ eigenstate corresponds to a projection on sublattice $B$ $(A)$ at the $K$ point whereas at the $K'$ point $|\psi_-\rangle$ $(|\psi_+\rangle)$ now corresponds to a particle in sublattice $A$ $(B)$.

\begin{center}
\begin{figure}[ht]
\vskip -0.4cm
\includegraphics[width=0.4\textwidth]{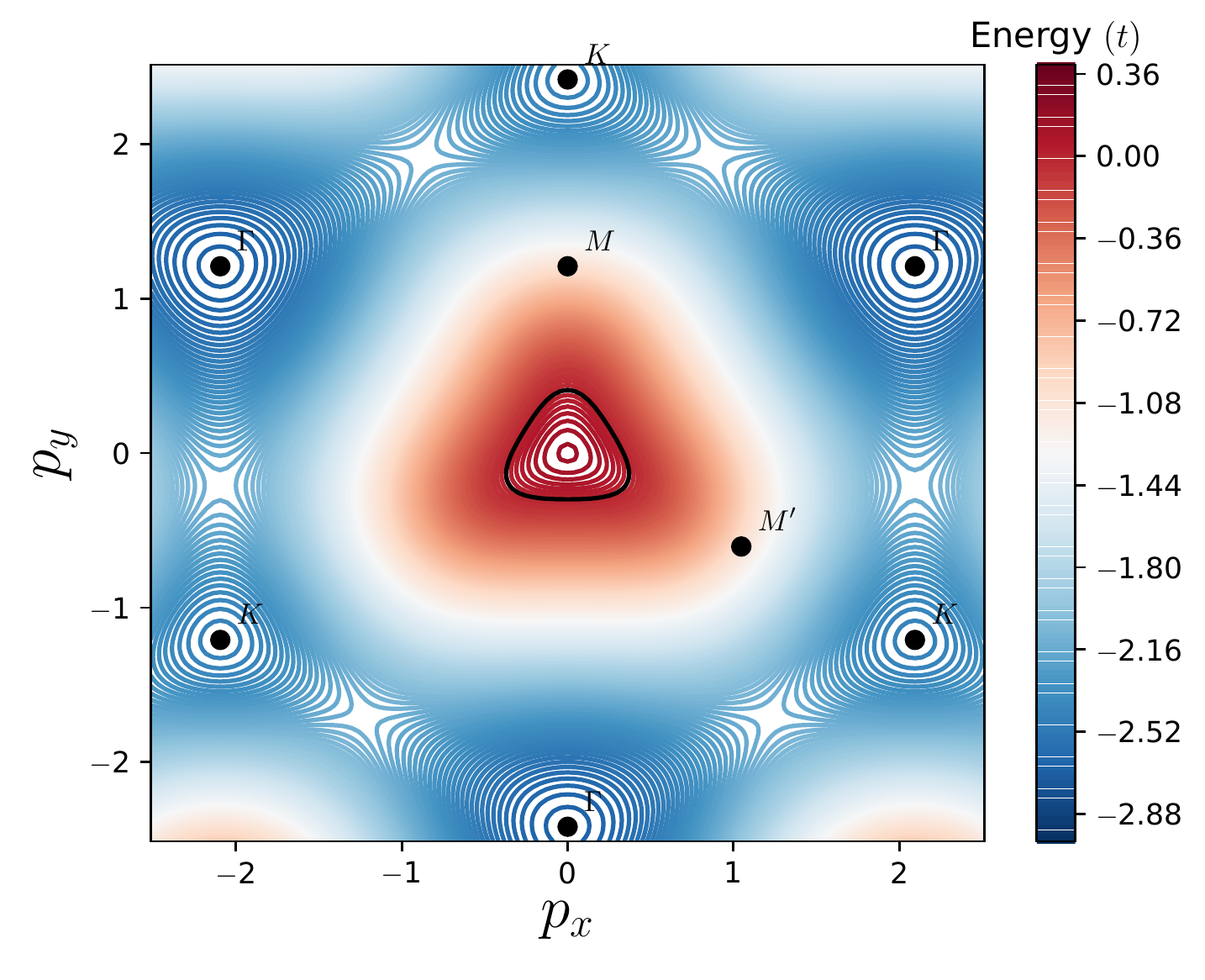}
\includegraphics[width=0.4\textwidth]{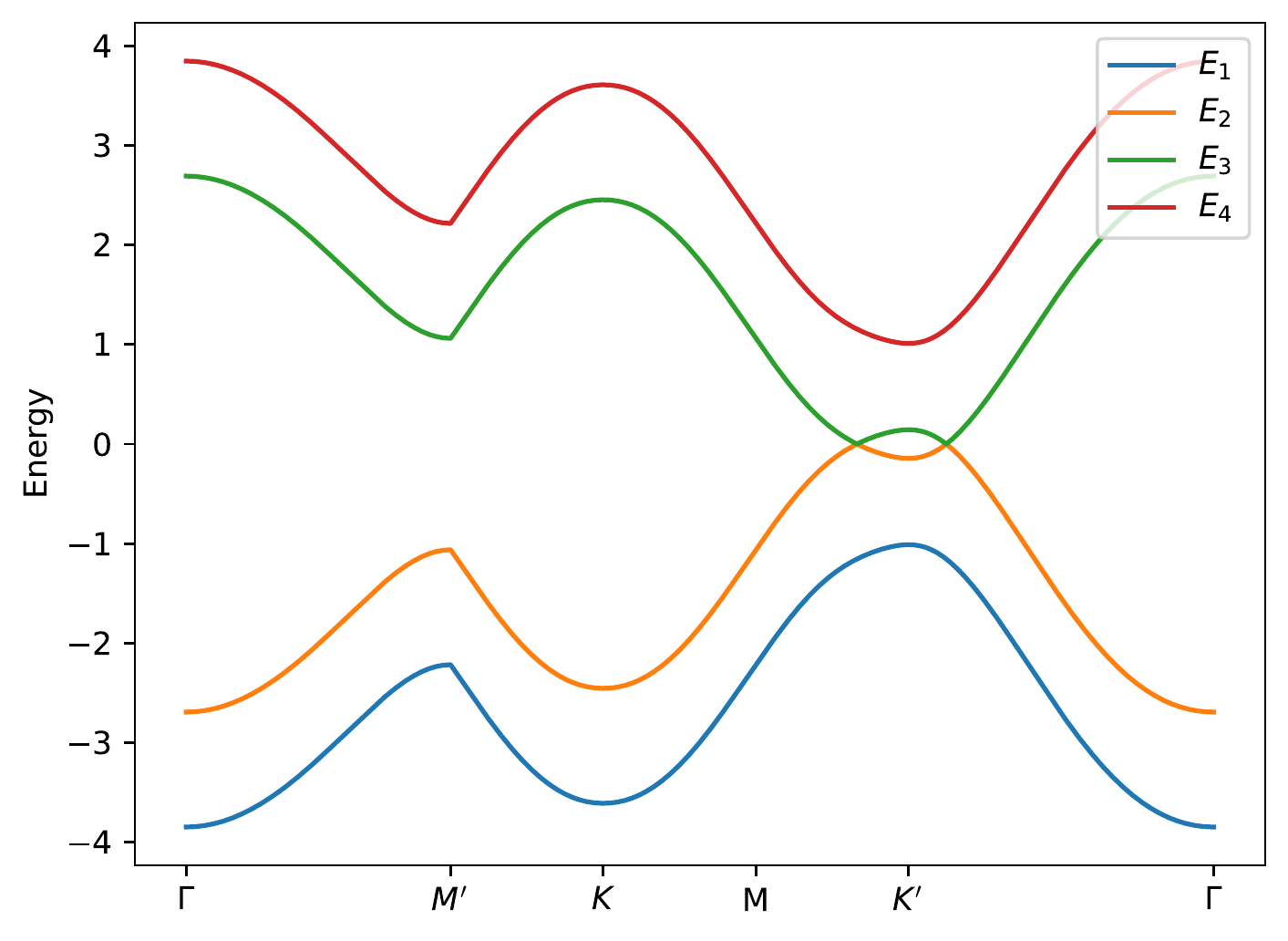}
\includegraphics[width=0.4\textwidth]{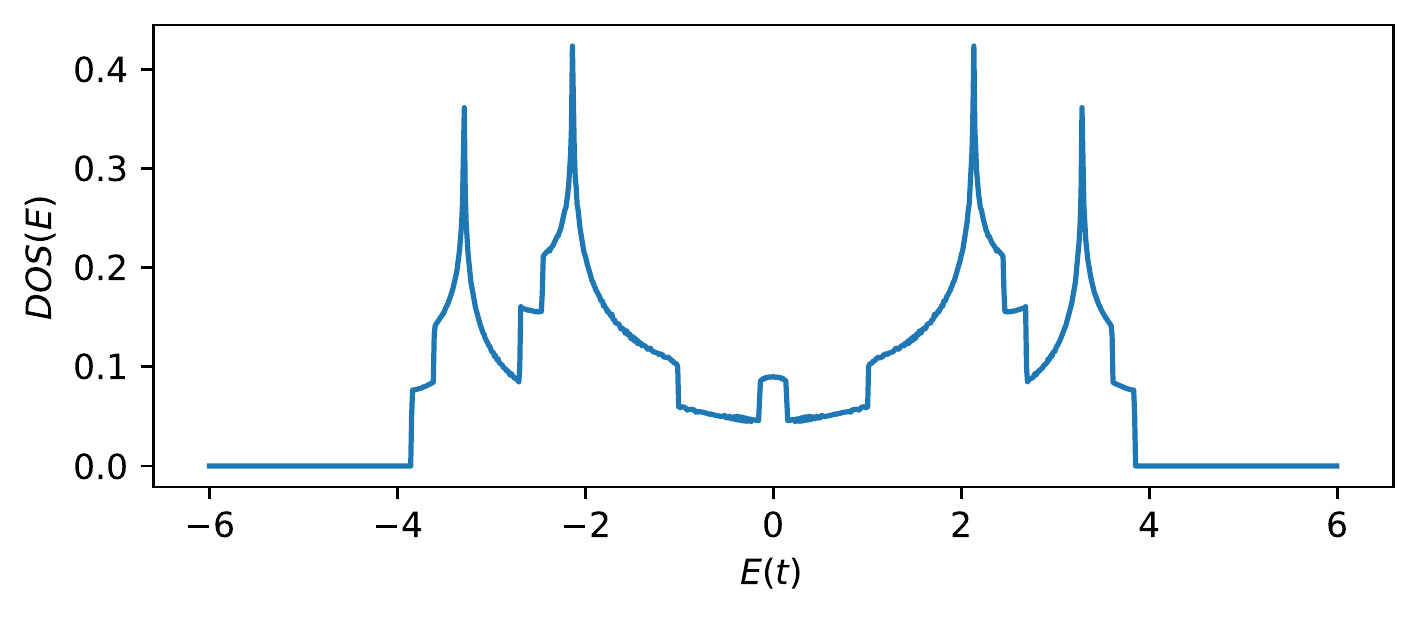}
\caption{(top) Contour representation of $E = r - \left|\bm{d}({\bf K}^\prime + \bm{p})\right|$ in the Brillouin zone centered around $K^\prime$ with the (triangular) nodal ring in black around $K^\prime$. (middle) Band structure of the system for a chosen path in the Brillouin zone. (bottom) Density of states of the system. The parameters chosen for all panels are $M = 3\sqrt{3}t/4$, $r = t/\sqrt{3}$, $t_2 = t/3$ with $a=1$.}
\label{Figures}
\end{figure}
\vskip -0.75cm
\end{center}

At half-filling, the ground state corresponds precisely to the two lowest bands being filled. 
At the $K$ Dirac point, the two-particles ground state corresponds to $|GS\rangle_K=|\psi_1(K)\rangle |\psi_2(K)\rangle$, which can be equivalently written developing the form of the ``radial'' eigenstates $|\psi_+\rangle$ and $|\psi_-\rangle$
at the pole of the sphere. This is equivalent to $|GS\rangle_K=e^{i\pi}c^{\dagger}_{B\uparrow} c^{\dagger}_{B\downarrow}|0\rangle$, producing two particles on the same sublattice. In second quantization, the identification reads $|\psi_1(K)\rangle = \frac{1}{\sqrt{2}}(-c^{\dagger}_{B\uparrow}+c^{\dagger}_{B\downarrow})|0\rangle$
and $|\psi_2(K)\rangle = \frac{1}{\sqrt{2}}(c^{\dagger}_{B\uparrow}+c^{\dagger}_{B\downarrow})|0\rangle$. Here, $|0\rangle$ refers to the vacuum at the $K$ point.
At the $K'$ Dirac point, the two-particles ground-state wavefunction then reads $|GS\rangle_{K'} =|\psi_1(K')\rangle |\psi_2(K')\rangle$ or equivalently 
$|GS\rangle_{K'} = \frac{1}{2}\left(c^{\dagger}_{A\uparrow}c^{\dagger}_{B\uparrow} - c^{\dagger}_{A\uparrow}c^{\dagger}_{B\downarrow}-c^{\dagger}_{A\downarrow}c^{\dagger}_{B\uparrow}+c^{\dagger}_{A\downarrow}c^{\dagger}_{B\downarrow}\right)|0\rangle$.
This form of entangled wavefunction, which also traduces the $\mathbb{Z}_2$ $\uparrow\leftrightarrow \downarrow$ symmetry, agrees with the structure of the eigenstates obtained when solving the $4\times 4$ energy matrix acting on $\psi({\bf k})=(c_{A{\bf k}\uparrow},c_{B{\bf k}\uparrow},c_{A{\bf k}\downarrow},c_{B{\bf k}\downarrow})$ and shows then a duality with the bilayer system studied in \cite{HH}. This analogy with the bilayer system and two entangled Bloch spheres \cite{HH} allows us to certify that, if Eq. (\ref{prerequisite}) is satisfied, then the graphene becomes a topological semimetal in the sense that it also acquires one chiral edge mode and has a quantized quantum Hall conductivity $\sigma_{xy}=C\frac{e^2}{h}$ with a total Chern number $C=\sum_{\alpha=\uparrow,\downarrow}|C_{\alpha}|=1$. The $\frac{1}{2}$ topological number per spin polarization can be viewed as a half Skyrmion, first introduced  as a possible solution of the Yang-Mills equation \cite{Alfaro}, and this also measures the entanglement properties at the $K'$ point \cite{KarynReview}. From the geometry on a sphere, the ground state is characterized by a total Berry phase $2\pi$ at the Dirac points \cite{HH} in accordance with the properties of band $1$ (in continuity with the limit $r=0=M$). The fact that $C_{\alpha}=\frac{1}{2}$ with $\alpha=\uparrow,\downarrow$ can also be understood from the fact that the lowest topological band $1$ is polarized along $|-\rangle_x$. The eigenstates of band $2$ at $K$ and $K'$ are similar to those in a non-topological charge density wave phase corresponding to a projection on $B$ sublattice.

Now, we address the Fermi-liquid properties of the topological nodal ring semimetal related to Figs. \ref{Figures}.
If we develop the energy spectrum of bands $2$ and $3$ close to $K'$ we observe a quadratic evolution characteristic of a Fermi `liquid' with both
electrons and holes quasiparticles. The free energy takes the usual form $F(T)=F(0)-\frac{\pi^2}{6}T^2 D(E_F)$ at a temperature $T$ with the surface-related density of states at the Fermi energy $D(E_F)=\frac{\tilde{d}_z-M}{\pi}\frac{1}{v_F^2}=\frac{m^*}{\pi}$ with $m^*$ corresponding to the mass of a quasiparticle in the sense of the kinetic term $\pm\frac{1}{2}v_F^2\frac{|{\bf p}|^2}{\tilde{d}_z-M}$ related to the curvatures of band $2$ and $3$ close to $K'$. Here, $\frac{D(E_F)}{2}$ refers to the density of states for band $2$ or $3$. 
Then, when $r > r_c^-$ where $r_c^- = \tilde d_z - M$, the system shows a finite density of states related to the apparition of the nodal ring. This traduces the entrance into the semimetallic topological phase. For $r < r_c^-$, the system goes into an insulating topological phase with $C=2$ associated to the contribution of the two lowest bands \cite{BilayerArticle}. The transition at $r=r_c^-$ is characterized by the radius of the nodal ring shrinking to zero from Eq. (\ref{radius}) and a jump in the (electronic) specific heat at fixed volume $\frac{\pi^2}{3}D(E_F)T$ which can be observed tuning the magnetic field $B_x$. The in-plane magnetic field produces a macroscopic magnetization along $x$ direction in the semimetal phase as the lowest band is polarized along $|-\rangle_x$ and band $2$ flips its spin polarization from $|+\rangle_x$ to $|-\rangle_x$ within the Brillouin zone from $K$ to $K'$. At $r<r_c^-$, band $2$ flips its spin polarization at $K'$, $|-\rangle_x\rightarrow |+\rangle_x$, such that the net magnetization along $x$ for the ground state disappears in agreement with the limit $r\rightarrow 0$. The semimetal develops a Pauli magnetic susceptibility $\chi\sim\mu_B^2 \frac{1}{\pi v_F^2}$ when tilting the magnetic field slightly along $z$ axis. This can be verified including a term $r_z\mathbb{I}\otimes s_z$ with $r_z=\mu_B B_z$ in Eq. (\ref{Hmodel}) and evaluating $\mu_B \langle s_z\rangle \frac{D(E_F)}{2}$ for band $3$ and band $2$ which show an opposite induced spin polarization. If we navigate between $r_c^-$ and $r_c^+=\tilde{d}_z+M$ in Eq. (\ref{prerequisite}) then the shape of the ring progressively evolves from a circle onto a triangle and close to $r_c^+$ we observe a similar circular nodal ring forming now around the $K$ point. The system also develops a gapless region around the $\Gamma$ point in the Brillouin zone. For $r>r_c^+$, the ground state is a `trivial'  (non-topological) insulator.

{\it Realization and Protection.---} Below, we elaborate on the realization of the topological phase in graphene and its stability towards disorder effects. From general grounds, the robustness of the band structure at the two crossing points for a fixed angle $\tilde{\varphi}$ resides in the presence of a ``double-orthogonality'' in the definition of the eigenstates related to the two crossing bands. To hybridize these two bands, this requires to flip simultaneously the pseudo-spin through the action of $\sigma^{\pm}$ and the spin through the action of $s_y\mp is_z$. Therefore, tilting the magnetic field $B_x$ in the plane through a perturbation of the form $\mathbb{I}\otimes s_y$ will not alter the semimetal and will simply rotate smoothly the spin eigenstates. Here, we emphasize that the operator $\sigma^{\pm}\otimes (s_y\mp is_z)$ that could couple or hybridize the two-crossing bands is not generated including higher orders in the development of the partition function, then protecting the band degeneracy. Within the implementation of Eq. (\ref{Hmodel}) in graphene, the term 
$\beta\sigma_z\otimes s_z$ which may also open a gap at the crossing points (the state $\sigma_z|\psi_-\rangle$ being not orthogonal to $|\psi_+\rangle$ and similarly for the spin sector) remains zero, protecting the topological nodal ring semimetal and the $\mathbb{Z}_2$ symmetry.

An important point to realize this model is the form of the term $M\sigma_z\otimes\mathbb{I}$. Suppose potentials $\{V_{\uparrow}^A, V_{\uparrow}^B, V_{\downarrow}^A, V_{\downarrow}^B\}$ resolved on a sublattice, then we have the general relation ${\cal \hat{V}}=V_{\uparrow}^A \hat{n}_{\uparrow}^A + V_\uparrow^B \hat{n}_{\uparrow}^B + V_\downarrow^A \hat{n}_{\downarrow}^A +V_{\downarrow}^B \hat{n}_{\downarrow}^B$ with
\begin{eqnarray}
{\cal \hat{V}}=\alpha\mathbb{I}\otimes s_z + M\sigma_z\otimes \mathbb{I}+\beta\sigma_z\otimes s_z,
\end{eqnarray}
and $\alpha=\frac{1}{4}(V_{\uparrow}^A +V_{\uparrow}^B - (V_\downarrow^A +V_\downarrow^B))$, $M=\frac{1}{4}((V_{\uparrow}^A-V_{\uparrow}^B) + (V_{\downarrow}^A-V_{\downarrow}^B))$ and  $\beta=\frac{1}{4}((V_{\uparrow}^A-V_{\uparrow}^B) - (V_{\downarrow}^A-V_{\downarrow}^B))$. Within the definitions of the Pauli matrices, we define $\hat{n}_j^i = \hat{P}_i\otimes \hat{P}_j$ with $\hat{P}_i$ the probability to occupy the sublattice $i$ and $\hat{P}_j$ the probability to observe the spin polarization $j$. Therefore, to realize a term $M\sigma_z\otimes \mathbb{I}$ in the Hamiltonian this requires to have $V_{\uparrow}^A = V_{\downarrow}^A=V_A$ and $V_{\uparrow}^B = V_{\downarrow}^B=V_B$ preserving the $\mathbb{Z}_2$ symmetry, implying then that $\beta=0=\alpha$ and $M=\frac{1}{2}(V_A-V_B)$. It is judicious to have global potentials $V_A$ and $V_B\neq V_A$ such that they keep a similar form in real space and momentum space. This can also be nano-engineered through a substrate in a charge density wave or Mott phase \cite{Mott}. A Coulomb interaction $V(n_{CDW}^A n_g^A + n_{CDW}^B n_g^B)$ with $\langle n_{CDW}^A \rangle=1$ and $\langle n_{CDW}^B\rangle=0$ representing here the mean density of (spin-polarized) particles on the substrate per sublattice produces an effective potential $V_{eff}=\frac{V}{2}(1+\sigma_z\otimes\mathbb{I})$. 

\begin{center}
\begin{figure}[ht]
\includegraphics[width=0.4\textwidth]{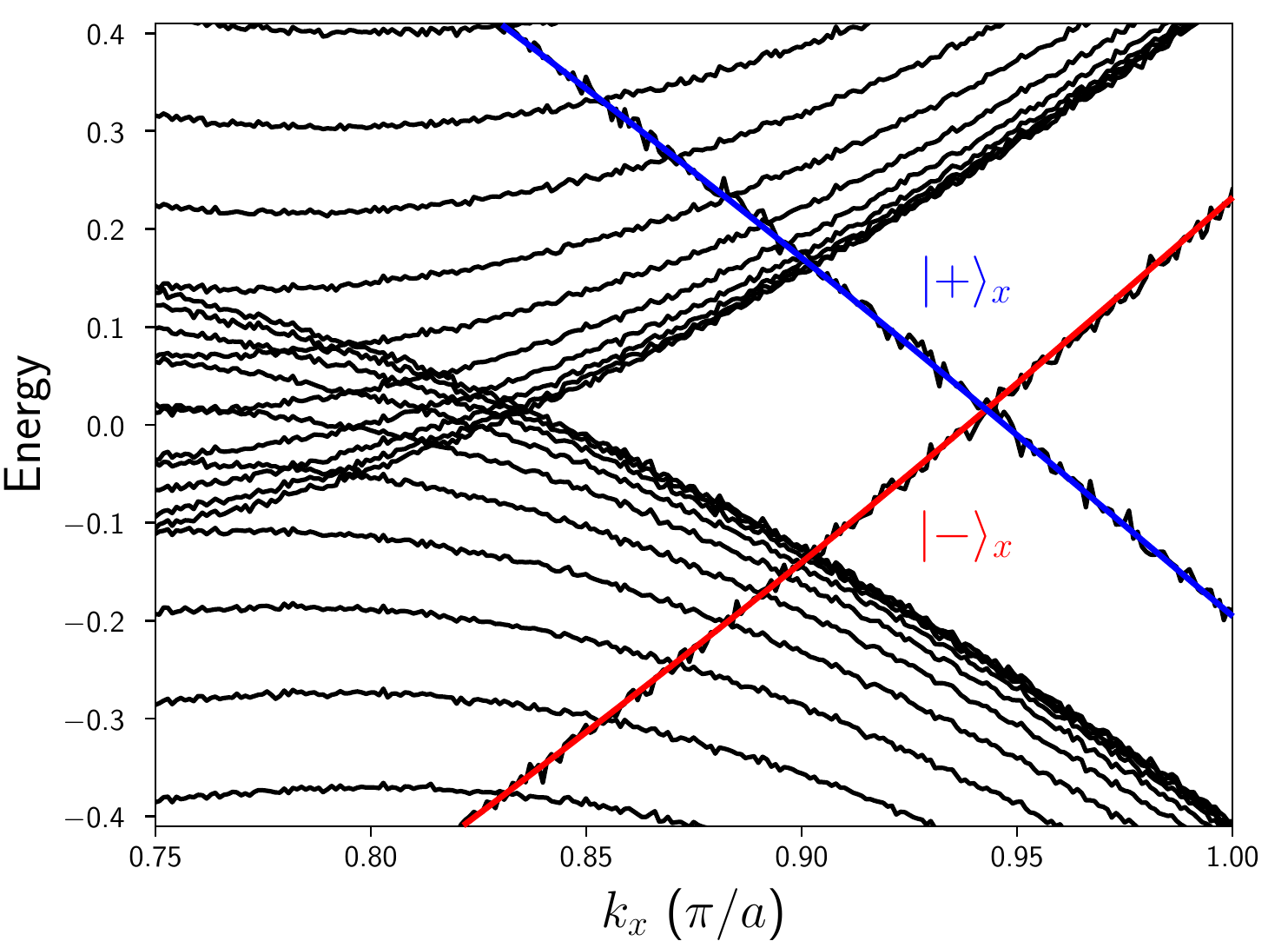}
\caption{Zoom on the band structure as a function of $k_x$ on the cylinder geometry at small disorder $U=0.03t$ showing the semimetal. Same parameters as in Fig. \ref{Figures}. The cylinder has $N=30$ sites in the non-periodic $y$ direction and one edge mode at the boundary
with the top/bottom disk: the red edge mode is linked with the lowest topological energy band $1$ and the blue edge mode is linked with band $4$.}
\vskip -1cm
\label{Figure2}
\end{figure}
\end{center}

The stability of the phase can be tested through various disorder profiles. Suppose a global disorder producing fluctuations of potential $\{\delta V_A;\delta V_B\}$ acting on sublattices $A$ and $B$ such that the chemical potential has small fluctuations around $\mu=0$.
In this case, the disorder can be inserted as a Gaussian random variable $\phi=\frac{1}{2}(\delta V_A-\delta V_B)$ essentially renormalizing the Semenoff mass term as $M\rightarrow M+\phi$.  We can then define the ensemble-averaged topological number \cite{Mott} as
$\langle C_{\alpha}\rangle = \int_{-\infty}^{+\infty} d\phi P(\phi)C_{\alpha}(\phi)$ with $P(\phi) = {\cal N} e^{-\frac{1}{2}\phi^2 \xi^{-1}}$ and ${\cal N}$ a normalization factor. As long as $\tilde{d}_z-M-\phi<r<\tilde{d}_z+M+\phi$ is satisfied then the topological number $C_{\alpha}(\phi)=1$ with $\alpha=\uparrow,\downarrow$ is stable and zero otherwise. Evaluating the integral then leads to the observation that the quantum Hall conductivity would only show exponentially small corrections as long as $\xi\ll \tilde{d}_z$ referring to the weak disorder limit. In Fig. \ref{Figure2}, we present a numerical evaluation of the band structure in a cylinder geometry with a disordered potential $V$ with each value $V(y)$ being a random variable taken uniformly in a window $[-U;U]$ where a partial Fourier transform is performed along the edge variable with $k=k_x$. The energy spectrum shows the survival of the edge mode (at each edge of the cylinder) until moderate ranges of $U$ compared to $\tilde{d}_z$. 

Since topological properties are characterized through the Dirac points \cite{KarynReview}, we describe below an implementation of the $\zeta \tilde{d}_z\sigma_z\otimes\mathbb{I}$ term resolved at $K$ and $K'$ shining the graphene lattice with circularly polarized light. The time-dependent Hamiltonian close to $K$ and $K'$ reads
\begin{eqnarray}
{\cal H}({\bf p},t) &=& M\sigma_z\otimes\mathbb{I}+v_F(p_x +e A_x(t))\sigma_x\otimes\mathbb{I} \\ \nonumber
&+& \zeta v_F(p_y+eA_y(t))\sigma_y\otimes\mathbb{I} +r\mathbb{I}\otimes s_x.
\end{eqnarray}
To induce a topological state both through a high-frequency Magnus approach \cite{Cayssol} in the Floquet theory and through the resonance in the rotating frame \cite{C2}, we introduce a circularly polarized vector potential of the form ${\bf A}=A_0(\sin(\omega t){\bf e}_x + \sin(\omega t+\frac{\pi}{2}){\bf e}_y)$ that would then produce simultaneously the presence of a left-handed and right-handed polarized waves interacting with one Dirac point respectively such that $\sigma_x A_x(t) +\zeta \sigma_y A_y(t) = + i A_0 \zeta \sigma^{+} e^{i\zeta \omega t} -  i A_0 \zeta \sigma^{-} e^{-i\zeta \omega t}$. Within the general Floquet formalism \cite{Cayssol}, then we verify that such a time-dependent perturbation indeed produces the required term $\tilde{d}_z=\frac{(e v_F A_0)^2}{\hbar\omega}$. Experiments in graphene have shown the possibility to generate such a topological term from circularly polarized light \cite{Hamburglight}. The measured quantum Hall conductivity depends on the light intensity at small $A_0$ and then can approach a quantized value when increasing $A_0$ \cite{AokiOka}. A similar situation may be reached through Floquet mechanism in cold atoms \cite{Jotzu,Hamburg} and light-matter systems \cite{Review1}.

The rotating frame approach is also useful to measure the topological number when driving from north to south pole on the sphere corresponding here to a pumping protocol of a charge from one Dirac point to the other. The rotating frame can be reached through the unitary transformation ${\cal U}_{\zeta}(t)=e^{-i\zeta\frac{\omega t}{2}\sigma_z}\otimes \mathbb{I}$ of angle $+\zeta \omega t$ \cite{C2}. This is equivalent to define the dressed eigenstates $|A'\rangle$ and $|B'\rangle$ such that $|B\rangle = e^{-i\zeta\frac{\omega t}{2}}|B'\rangle$ and $|A\rangle = e^{i\zeta\frac{\omega t}{2}}|A'\rangle$. The Hamiltonian becomes
\begin{equation}
{\cal H}_{eff}({\bf p}) = {\cal U}_{\zeta} {\cal H}({\bf p},t) {\cal U}_{\zeta}^{-1} + \zeta \tilde{d}_z\sigma_z\otimes \mathbb{I}
\end{equation}
with $\tilde{d}_z=\frac{\hbar\omega}{2}$ at the $K$ point. A time-independent Hamiltonian can be reached if we adjust the angle $\tilde\varphi=-\omega t$. Suppose for this protocol that $A_0\rightarrow 0$, then in the $\{|A'\rangle; |B'\rangle\}$ basis we obtain the Hamiltonian 
\begin{equation}
{\cal U}_{\zeta}{\cal H}{\cal U}_{\zeta}^{-1} = (\zeta \tilde{d}_z +M)\sigma_z\otimes \mathbb{I} + v_F|{\bf p}|\sigma_x \otimes \mathbb{I} + r\mathbb{I}\otimes s_x.
\end{equation}
This corresponds to the required Hamiltonian in Eq. (\ref{Hmodel}) if we navigate effectively along the line corresponding to an azimuthal angle $\varphi_{eff}=0$ on the Bloch sphere. To observe the topological invariant this requires to measure the electric dipole polarization on the lattice at the specific wave-vectors ${\bf K}$ and ${\bf K}'$ related to the two-dimensional lattice.  In this way, it seems possible to detect $C=1$ through the electrical polarization $\langle \sigma_{jz}\rangle$ resolved at the specific $K$ and $K'$ Dirac points within the Brillouin zone from the rotating frame adjusting $\tilde{d}_z-M<r<\tilde{d}_z+M$ with $r=-\frac{\gamma B_x}{2}$. 

{\it Stability towards Interaction Effects.---} The stability of the nodal ring semimetal band structure towards interaction effects can be understood from renormalization group arguments for two-dimensional (circular) Fermi surfaces \cite{Schulz}. In particular, the Pauli principle does not allow for inter-band interaction effects between spin-polarized states $|-\rangle_x$ and $|+\rangle_x$ associated to bands $2$ and $3$. This can also be understood through a (variational) mean-field approach \cite{stochastic} which reflects on the importance of symmetries at the band crossing points to preserve the stability of the nodal ring semimetal. A Hubbard interaction in real space, at site $i$, $Uc^{\dagger}_{i\uparrow}c_{i\uparrow}c^{\dagger}_{i\downarrow}c_{i\downarrow}$ can be equivalently written through mean-field variables $\phi_r = -\frac{1}{2}\langle S_r\rangle$ associated to the spin observables ${\bf S}=c_i^{\dagger}{\bf s}c_i$ and can be incorporated within the matrix approach in momentum space \cite{stochastic}. From the Hamiltonian (\ref{Hmodel}), we have $\phi_y=0$. Similarly, $\phi_z=0$ since an effective coupling hybridizing bands $2$ and $3$ at the crossing points would necessitate an operator of the form $\sigma^+\otimes(s_y-is_z)$ or $\sigma^-\otimes(s_y+is_z)$. The fact that $\phi_z=0$ is also in agreement with the longitudinal spin susceptibility being a function of $v_F$ only. Therefore, interactions simply renormalize the term $r$ as $\tilde{r}=(r+U\phi_x)$. For the filled energy band $2$, around the $K'$ Dirac point we have $\phi_x=+\frac{1}{2}$ which indeed ensures the stability of the semimetal as long as $\tilde{d}_z-M<r+\frac{U}{2}<\tilde{d}_z+M$. Developing the energy of band $2$ close to $K'$ in the presence of the $U\phi_x$ channel, the mass $m^*$ (and therefore the density of states) of a (quasi-)particle also remains identical. 

{\it Conclusion.---} To summarize, we have introduced a protected topological nodal ring semimetal in graphene showing the characteristics of a Fermi liquid with $\mathbb{Z}_2$ symmetry in the bulk coexisting with a robust chiral edge mode and giving rise to a quantum anomalous Hall response. This system offers further perspectives on the possibility to realize topological semimetals with the equivalent of matter and antimatter in the bulk and a chiral edge mode.  This may also lead to developments related to quantum materials, cold atoms and light-matter systems. Related to this quest, it is interesting to mention here the very recent observation of a quantized (thermoelectric) Hall plateau in graphite as a nodal line semimetal \cite{Tokyo}. 

This work is funded by the Deutsche Forschungsgemeinschaft (DFG, German Research Foundation) under Project No.  277974659 via Research Unit FOR 2414 and via ANR BOCA. S.A.S. is grateful to Ecole Polytechnique regarding the funding of his PhD thesis. K.L.H. also acknowledges interesting discussions related to this work at Aspen Center for Physics, which is supported by National ScienceFoundation grant PHY-1601671, and at Dresden conference TOPCOR22.

\end{document}